\UseRawInputEncoding
\documentclass[]{interact}

\usepackage{epstopdf}
\usepackage[caption=false]{subfig}

\usepackage[longnamesfirst,sort]{natbib}
\bibpunct[, ]{(}{)}{;}{a}{,}{,}
\renewcommand\bibfont{\fontsize{10}{12}\selectfont}

\theoremstyle{plain}

\theoremstyle{definition}

\theoremstyle{remark}

\newtheorem{Theorem}{Theorem}[section]

\DeclareMathOperator*{\argmin}{argmin}
\def\x{\mathrm{x}}
\def\y{\text{y}}

\begin{document}


\title{Model Averaging for Generalized Linear Models in
Fragmentary Data Prediction}

\author{
\name{Chaoxia Yuan\thanks{CONTACT Chaoxia Yuan. Email: chaoxiayuan@163.com}, Yang Wu and Fang Fang}
\affil{KLATASDS - MOE, School of Statistics, East China Normal University, Shanghai 200062, P. R. China}
}

\maketitle

\begin{abstract}
Fragmentary data is becoming more and more popular in many areas which brings big challenges to researchers and data analysts. Most existing methods dealing with fragmentary data consider a continuous response while in many applications the response variable is discrete. In this paper we propose a model averaging method for generalized linear models in fragmentary data prediction. The candidate models are fitted based on different combinations of covariate availability and sample size. The optimal weight is selected by minimizing the Kullback-Leibler loss in the completed cases and its asymptotic optimality is established. Empirical evidences from a simulation study and a real data analysis about Alzheimer disease are presented.
\end{abstract}


\section{Introduction}\label{sec1}

Our study is motivated by the study of Alzheimer disease (AD). Its main clinical features are the decline of cognitive function, mental symptoms, behavior disorders, and the gradual decline of activities of daily living.  It is the most common cause of dementia among people over age of 65, yet no prevention methods or cures have been discovered. The Alzheimer¡¯s Disease Neuroimaging Initiative (ADNI, {\it http://adni.loni.usc.edu}) is a global research program that actively supports the investigation and development of treatments that slow or stop the progression of AD. The researchers collect multiple sources of data from voluntary subjects including: cerebrospinal fluid (CSF), positron emission tomography (PET), magnetic resonance imaging (MRI) and genetics data (GENE). In addition,  mini-mental state examination (MMSE) score is collected for each subject, which is an important diagnostic criterion for AD. Our target is to establish a model focusing on the AD prediction (Alzheimer or not). This task is relatively easy if the data are fully observed. However, in ADNI data, not all the data sources are available for each subject. As we can see from Table \ref{table1} in Section \ref{sec5}, among the total of 1170 subjects, only 409 of them have all the covariate data available, 368 of them do not have the GENE data, 40 of them do not have the MRI data, and so on. Such kind of ``fragmentary data" nowadays is very common
in the area of medical studies, risk management, marketing research and social
sciences (Fang et al., 2019; Zhang et al., 2020b; Xue and Qu, 2021; Lin et al., 2021). But the extremely high missing rate and complicated
missing patterns bring big challenges to the analysis of fragmentary data.

In this paper we discuss the model averaging methods for fragmentary data prediction. Model averaging is historically proposed as an alternative to model selection. The most well-known model selection methods include AIC (Akaike, 1970), Mallows $C_p$ (Mallows, 1973),  BIC (Schwarz, 1978), lasso (Tibshirani, 1996), smoothly clipped absolute deviation (Fan and Li, 2001), sure independence screening (Fan and Lv, 2008) and so on.

Model averaging, unlike most variable selection methods which focus on identifying a single ``correct model",
aims to the prediction accuracy given several predictors (Ando and Li, 2014). Without ``putting all inferential eggs in one unevenly woven basket" (Longford, 2005), model averaging takes all the candidate models into account and makes prediction by a weighted average, which can be classified into Bayesian and frequentist model averaging. In this paper, we focus on frequentist model averaging (Buckland et al., 1997; Yang, 2001, 2003; Hjort and Claeskens, 2003; Leung and Barron, 2006; Hansen 2007, among many others) and refer readers being interested in Bayesian model averaging to Hoeting et al. (1999) and the references therein.  Researchers have developed many frequestist model averaging methods over the past two decades. To just name a few, the smoothed AIC and smoothed BIC (Buckland et al., 1997), Mallows model averaging (Hansen, 2007), Jackknife model averaging (Hansen and Racine, 2012) and heteroskedasticity-robust $C_p$ (Liu and Okui, 2013) mainly focus on low dimensional linear models.  Ando and Li (2014) and Zhang et al. (2020a) consider least squares model averaging with high dimensional data. For more complex models, we have model averaging for generalized linear models (Zhang et al., 2016; Ando and Li, 2017), quantile regression (Lu and Su, 2015), semiparametric ``model averaging marginal regression" for time series (Li et al., 2015; Chen et al., 2018), model averaging for covariance matrix estimation (Zheng et al., 2017), varying-coefficient models (Li et al., 2018; Zhu et al., 2019), vector autoregressions (Liao et al., 2019), semiparametric model averaging for the dichotomous response (Fang et al., 2020), and so on.

All the model averaging methods mentioned above assume that the data are fully observed and can not be applied to fragmentary data directly. Due to the extra large missing rate and complex response patterns, the traditional missing data techniques such as imputation and inverse propensity weighting (Little and Rubin, 2002; Kim and Shao, 2013) can not be efficiently applied either. Recently, Zhang et al. (2020b), Xue and Qu (2021) and Lin et al. (2021) develop some methods for block-wise missing or individual-specific missing data. But they only consider a continuous response.

On the other hand, Schomaker et al. (2010) and Dardanoni et al. (2011, 2015) propose model averaging methods based on imputation with no asymptotic optimality. Zhang (2013) proposes a model averaging method by impute the missing data by zeros for linear models. Liu and Zheng (2020) extends it to generalized linear models. In the context of fragmentary data, Fang et al. (2019) proposes a model averaging method to select weight by cross-validation on the complete cases and shows its advantage to the previous model averaging methods. Ding et al. (2021) extends it to multiple quantile regression. Asymptotic optimalities are established for the last four methods but they are only applicable to a continuous response except Liu and Zheng (2020).

In this paper, we propose a model averaging method for fragmentary data prediction in generalized linear models. The candidate models are fitted based on different combinations of covariate availability and sample size. The optimal weight is selected by minimizing the Kullback-Leibler loss in the completed cases and its asymptotic optimality is established. Unlike the methods in Fang et al. (2019), our method does not need to refit the candidate models in the complete cases for weight selection. Empirical results from a simulation study and a real data analysis about Alzheimer disease show the superiority of the proposed method.

The paper is organized as follows. Section \ref{sec2} discusses the proposed method in details. Asymptotic optimality is established in Section \ref{sec3}. Empirical results of a simulation study and a real data analysis are presented in Section \ref{sec4} and Section \ref{sec5}, respectively. Section \ref{sec6} concludes the paper with some remarks. All the proofs are provided in the Appendix.

 \section{The Proposed Method}\label{sec2}

For illustration, we consider the fragmentary data in Fang et al. (2019)  as presented in Table 1. A random sample consists of $n$ subjects with a response variable $Y$ and a covariate  set $D = \{X_{j},j = 1,\cdots,p\}$. Each subject $i$ only has covariate data available for a subset $D_{i}\subseteq D$. In Table 1, $D_1=D_2=\{X_1,\cdots,X_8\}$, $D_3=\{X_1,X_2,X_3\}$ and so on.
Let $\left\lbrace \Delta_{k},k=1,\cdots,K\right\rbrace$ be the set of all different response patterns for $ D_i, i = 1, \cdots, n$. Note that $1\leq K\leq 2^p-1$ if we assume $D_i\neq \Phi$ for any $i$, where $\Phi$ presents the empty set, but $K$ may be much smaller than $2^p-1$. In Table 1, $2^p-1=255$ but $K=7$. Note $\Delta_1=\{X_1,\cdots,X_8\}$, $\Delta_2=\{X_1,X_2,X_3\}$, $\cdots$, and $\Delta_7=\{X_1,X_2,X_7,X_8\}$. For notation simplicity, throughout the paper we also use $D_i$ or $\Delta_k$ to denote the set of indices of the covariates in $D_i$ or $\Delta_k$, e.g., $D_i=\{X_1,X_2,X_3\}=\{1,2,3\}$ or $\Delta_k=\{X_1,X_4,X_5,X_6\}=\{1,4,5,6\}$.

Denote $T_k = \{i\colon D_i = \Delta_{k}\}$ as the subject set with the response pattern $\Delta_{k}$. Then
$\{1,2,\cdots,n\} =\bigcup_{k=1}^{K} T_{k}$ and $T_k\bigcap T_{l}=\Phi$ if $k \neq l$. Assume that the subjects have been rearranged so that $\max_{i\in T_k}i=\min_{i\in T_{k+1}}i-1$. Denote $S_k = \{i : D_i \supseteq \Delta_k\} $ as the subject set with covariates in $\Delta_k$ being available. In Table 1, $T_1=\{1,2\}$, $S_1=\{1,2\}$, $T_2=\{3\}$, $S_2=\{1,2,3,4\}$, $\cdots$, $T_7=\{9,10\}$ and $S_7=\{1,2,4,9,10\}$.

\begin{table}[h]
\centering
\tbl{An illustrative example for fragmentary data.}
{\begin{tabular}{c|c|cccccccc}
    \hline
    Subject & $Y$     & $X_1$  & $X_2$  & $X_3$  & $X_4$  & $X_5$  & $X_6$  & $X_7$  & $X_8$ \\
    \hline
    1     & $\ast$     & $\ast$     & $\ast$     & $\ast$     & $\ast$     & $\ast$     & $\ast$     & $\ast$     & $\ast$ \\
    2     & $\ast$     & $\ast$     & $\ast$     & $\ast$     & $\ast$     & $\ast$     & $\ast$     & $\ast$     & $\ast$ \\
    3     & $\ast$     & $\ast$     & $\ast$     & $\ast$     &       &       &       &       &  \\
    4     & $\ast$     & $\ast$     & $\ast$     & $\ast$     &       &       &       & $\ast$     & $\ast$ \\
    5     & $\ast$     & $\ast$     &       &       & $\ast$     & $\ast$     & $\ast$     &       &  \\
    6     & $\ast$     & $\ast$     &       &       & $\ast$     & $\ast$     & $\ast$     &       &  \\
    7     & $\ast$     & $\ast$     &       &       &       &       &       &       &  \\
    8     & $\ast$     & $\ast$     &       &       &       & $\ast$     &       & $\ast$     & $\ast$ \\
    9     & $\ast$     & $\ast$     & $\ast$     &       &       &       &       & $\ast$     & $\ast$ \\
    10    & $\ast$     & $\ast$     & $\ast$     &       &       &       &       & $\ast$     & $\ast$ \\
    \hline
    \end{tabular}}
    \tabnote{{$\ast$}the datum is available.}
\label{tab:addlabel}%
\end{table}

Our target is to make prediction given the fragmentary data $\{(y_i, x_{ij}), i = 1,\cdots,n,j\in D_i\}$, where $y_i$ and $x_{ij}$ are observations of $Y$ and $X_j$ whenever they are observed. Specifically,
consider that  $Y$ given $X=(X_1,\cdots,X_p)^T$ has an exponential family distribution
\begin{align}\label{tm}
f(Y|X)=\exp\left\{\frac{Y\theta(X)-b(\theta(X))}{\phi}+c(Y,\phi)\right\}
\end{align}
for some known functions $b(\cdot)$, $c(\cdot,\cdot)$ and a known dispersion parameter $\phi$. The canonical parameter $\theta(\cdot)$ is unknown. For a new subject with available covariate data $D^*\in\{\Delta_k,k=1,\cdots,K\}$, we need to estimate $\theta(D^*)$.

Without loss of generality, we assume that $\Delta_1 = D = \{1, 2, \cdots , p\}$. Then $S_1 = T_1$ is the CC (complete cases) sample in the missing data terminology. Similar to Fang et al. (2019), we mainly focus on prediction of $\theta(\x^\ast)$ with $\x^\ast= (x_{1}^{\ast}, \cdots , x_{p}^{\ast})^T$ from pattern $\Delta_1$, i.e., $D^*=\Delta_1=D$. Any $\x^\ast$ from other pattern $\Delta_{k}$ can be handled in the same way by ignoring the covariates not in $\Delta_{k}$, which will be illustrated in the real data analysis.

As discussed in Fang et al. (2019), there exists a natural trade-off between the covariates included in the prediction model and available sample size. Taking Table 1 as an example, if we want to include all the 8 covariates in the model, only subjects 1 and 2 can be used without imputation. But if we only include the first covariate in the model, all the 10 subjects can be used. This trade-off naturally prepares a sequence of candidate models for model averaging.

  Specifically, we can fit a generalized linear model $M_k$ on the data $\{(y_{i},x_{ij}),i\in S_k,j\in \Delta_k\}$ and try to combine the prediction results from all the candidate models $\{M_k,k=1,\cdots,K\}$. Denote $\text{y}=(y_1,\cdots,y_n)^T$ and $x_i=(x_{i1},\cdots,x_{ip})^T$. Besides, the design matrix of $M_k$ is expressed as $\text{X}_k=(x_{ij}: i\in S_k,j\in \Delta_k)\in \mathbb{R}^{n_k\times p_k}$, where $n_k=|S_k|$ and $p_k=|\Delta_k|$. We assume $n_1\geq p$. Consequently $n_k\geq p_k$ since $n_k\geq n_1$ and $p_k\leq p$.
The candidate model $M_k$ is expressed as
\begin{equation} f(y_i|\theta_i^{(k)},\phi)=\exp\left\{\frac{y_i\theta_i^{(k)}-b(\theta_i^{(k)})}{\phi}+c(y_i,\phi)\right\},i\in S_k,\label{Mk}
\end{equation}
where $\theta_i^{(k)}$ is the $i^{\text{th}}$ element of the parameter $\theta_{(k)}=(\theta_i^{(k)},i\in S_k)^T$. It is modeled by a linear model $\theta_{(k)}=\text{X}_k\beta_{(k)}$.
Denote the maximum likelihood estimator of $\beta_{(k)}$ by $\hat\beta_{(k)}$. Note that we do not assume that the true model $\theta(X)$ in (\ref{tm}) is indeed a linear function of $X$. Thus,
all the candidate models can be misspecified.
For a new $\x^\ast=(x_1^\ast,\cdots,x_p^\ast)^T$, we predict $\theta(\x^\ast)$ by
$$\hat\theta^\ast(w)=\sum_{k=1}^Kw_k\x_{k}^{\ast^T}\hat\beta_{(k)}= \x^{\ast^T}\hat\beta(w),$$
 where $\x_{k}^{\ast}=(x_j^{\ast}:j\in \Delta_k)^T$, $\hat\beta(w)=\sum_{k=1}^Kw_k\Pi^T_k\hat\beta_{(k)}$, $\Pi_k$ is a projection matrix of size $p_k\times p$ consisting of 0 or 1 such that $\x_k^{\ast}=\Pi_k\x^{\ast}$, and the weight vector $w=(w_1,\cdots,w_K)^T$ belongs to
$$\mathcal{H}_n=\left\{w\in [0,1]^K:\sum_{k=1}^{K}w_k=1\right\}.$$

 Let $\theta\{\hat\beta(w)\}=(\theta_1\{\hat\beta(w)\},\cdots,\theta_{n_1}\{\hat\beta(w)\})^T=\text{X}_1\hat\beta(w)$ be the model averaging estimator of $\theta_{(1)}$.
 Our weight choice criterion is motivated by the Kullback-Leibler (KL) loss in Zhang et al. (2016) and is defined as
follows. Denote the true value of $\theta_{(1)}$ as $\theta_0=(\theta_{0_{1}},\cdots,\theta_{0_{n_1}})^T$. Let $\y^*=(y^*_1,\cdots,y^*_{n_1})^T$ be another realization from $f(\cdot|\theta_0,\phi)$ and independent of $\y$. The KL loss of $\theta\{\hat\beta(w)\}$ is
\begin{align}\label{KL}
  \text{KL}(w)&=2\sum_{i\in S_1}E_{\y^*}\left\{\log\{f(\y^*_i|\theta_{0i},\phi)\}-\log(f[\y^*_i|\theta_i\{\hat{\beta}(w)\},\phi])\right\}\nonumber\\ &=2\phi^{-1}B\{\hat\beta(w)\}-2\phi^{-1}\mu_{S_1}^T\theta\{\hat\beta(w)\}-2\phi^{-1}B_0+2\phi^{-1}\mu_{S_1}^T\theta_0\nonumber\\
  &=2J(w)-2\phi^{-1}B_0+2\phi^{-1}\mu_{S_1}^T\theta_0,
\end{align}
where $B_0=\sum_{i\in S_1}b(\theta_{0i})$, $B\{\hat\beta(w)\}=\sum_{i\in S_1}b[\theta_i\{\hat\beta(w)\}]$, $\mu_{S_1}=(\mu_{S_1,1},\cdots,\mu_{S_1,n_1})^T=(E(y_i|i\in T_1),i=1,\cdots,n_1)^T$ and
\begin{align}
J(w)&=\phi^{-1}B\{\hat\beta(w)\}-\phi^{-1}\mu^T_{S_1}\theta\{\hat\beta(w)\}\nonumber\\
      &=\phi^{-1}\left\{\sum_{i \in S_1}b[\theta_i\{\hat{\beta}(w)\}]-\sum_{i \in S_1}\mu_{S_1,i}\theta_i\{\hat{\beta}(w)\}\right\}.\nonumber
\end{align}

As Zhang et al. (2016) discussed, we would obtain a weight vector by minimizing $J(w)$ given $\mu_{S_1}$.  However, it is infeasible in practice to do so since the parameter $\mu_{S_1}$ is unknown. Instead, we replace $\mu_{S_1}$ by $y_{S_1}=(y_i,i\in S_1)^T$ and add an penalty term to $J(w)$ to avoid overfitting, which gives us the following weight choice criterion
\begin{align}
\mathcal{G}(w)=2\phi^{-1}\left\{\sum_{i\in S_1}b[\theta_i\{\hat{\beta}(w)\}]-\sum_{i \in S_1}y_i\theta_i\{\hat{\beta}(w)\}\right\}+\lambda_n\sum_{k=1}^Kw_kp_k,\nonumber
\end{align}
where $\lambda_n\sum_{k=1}^Kw_kp_k$ is the penalty term, and $\lambda_n$ is a tuning parameter that usually takes value 2 or $\log(n_1)$. The optimal weight vector is defined as
\begin{align}\label{KL1}
\hat{w}=\argmin_{w\in \mathcal{H}_n}\mathcal{G}(w).
\end{align}

\vspace{0.4cm}
\textbf{Remark 1:} Basically, our idea is to use all available data to estimate parameters for each candidate model and use CC data to construct the optimal weights. This is similar to Fang et al. (2019) that deals with linear models for fragmentary data. However, unlike Fang et al. (2019), our proposed method does not need to refit the candidate models in the CC data to decide the optimal weight. Similar to Zhang (2013), Liu and Zheng (2020) selects weights by applying KL loss to the entire data with unavailable covariate data replaced by zeros, which does not perform quite well in the empirical studies.
\vspace{0.4cm}

\textbf{Remark 2:} Under the logistic regression model, $\phi=1$ and $b(\theta)=\log(1+e^\theta)$. Let $\theta_i\{\hat{\beta}(w)\}=\log\frac{\hat{p}_i(w)}{1-\hat{p}_i(w)}$. Then
 \begin{align}
J(w)&=\sum_{i\in S_1}\log\left[1+e^{\theta_i\{\hat{\beta}(w)\}}\right]-\sum_{i \in S_1}\mu_{S_1,i}\theta_i\{\hat{\beta}(w)\}\nonumber\\
&=-\sum_{i\in S_1}\log\left\{1-\hat{p}_i(w)\right\}-\sum_{i\in S_1}\mu_{S_1,i}\log\frac{\hat{p}_i(w)}{1-\hat{p}_i(w)}\nonumber
\end{align}
\begin{align}\label{KL2}
&=-\sum_{i\in S_1}\big[\mu_{S_1,i}\log\hat{p}_i(w)+(1-\mu_{S_1,i})\log\{1-\hat{p}_i(w)\}\big]
\end{align}
and
\begin{align}
\mathcal{G}(w)=-2\sum_{i\in S_1}\big[y_i\log\hat{p}_i(w)+(1-y_i)\log\{1-\hat{p}_i(w)\}\big]+\lambda_n\sum_{k=1}^Kw_kp_k.\nonumber
\end{align}

\section{Asymptotic Optimality}\label{sec3}

Let $\beta^*_{(k)}$ be the parameter vector that minimizes the KL divergence between the true model and the $k^{\text{th}}$ candidate model (\ref{Mk}). From Theorem 3.2 of White (1982), we know that, under certain regularity conditions,
\begin{align}\label{con}
\hat{\beta}_{(k)}-\beta^*_{(k)}=O_p(n_k^{-1/2})=O_p(n_1^{-1/2}).
\end{align}

Let $\epsilon_{S_1}=(\epsilon_{S_1,1},\cdots,\epsilon_{S_1,n_{1}})^T=(y_1,\cdots,y_{n_1})^T-\mu_{S_1}$, $\bar{\sigma}^2=\max_{i\in S_1}\text{Var}(\epsilon_{S_1,i})$, $\beta^*(w)=\sum_{k=1}^Kw_k\Pi_k^T\beta^*_{(k)}$,
$$\text{KL}^*(w)=2\phi^{-1}B\{\beta^*(w)\}-2\phi^{-1}B_0-2\phi^{-1}\mu^T_{S_1}[\theta\{\beta^*(w)\}-\theta_0],$$
and $\xi_n=\inf_{w\in\mathcal{H}_n}\text{KL}^*(w)$. We assume the following conditions.

\vspace{0.5cm}
\textbf{(C1)}: $\parallel\text{X}^T_1\mu_{S_1}\parallel=O(n_1),\parallel\text{X}^T_1\epsilon_{S_1}\parallel=O_p(n_1^{1/2})$, and uniformly for $w\in \mathcal{H}_{n}$,
$$\parallel\partial B(\beta)/\partial \beta^T|_{\beta=\tilde{\beta}(w)}\parallel=O_{p}(n_1)$$
for every $\tilde{\beta}(w)$ between $\hat{\beta}(w)$ and $\beta^*(w)$.

\textbf{(C2)}: Uniformly for $k\in\{1,\cdots,K\}$, $n_1^{-1}\bar{\sigma}^2\parallel\theta(\Pi_k^T\beta^*_{(k)})\parallel^2=O(1)$.

\textbf{(C3)}: $n_1\xi_n^{-2}=o(1)$.

\vspace{0.5cm}

The following theorem establishes the asymptotic optimality of the model averaging estimator $\theta\{\hat{\beta}(\hat{w})\}$.
{\Theorem \label{th-1}
Under equation (\ref{con}), conditions (C1)$\sim$(C3), and $n_1^{-1/2}\lambda_n=O(1)$, we have
\begin{equation}\label{}
  \frac{\text{KL}(\hat w)}{\inf_{w\in \mathcal{H}_n}\text{KL}(w)}\rightarrow_p1,\nonumber
\end{equation}
where $\text{KL}(w)$ is defined in (\ref{KL}) and $\hat w$ is defined in (\ref{KL1}). \\
}

The Conditions (C1)-(C3) are similar to Conditions (C.1)-(C.3) in Zhang et al. (2016). What is slightly different is the order $O(n_1)$ other than $O(n)$. It is rational because our weights selection is based on CC ($i\in S_1$) data with sample size $n_1$. Condition (C3) requires that $\xi_n$ grows at a rate no slower than $n_1^{1/2}$, which is the same as the third part of Condition (A7) of Zhang et al. (2014), and is also implied by Conditions (7) and (8) of Ando
and Li (2014). Condition (C3) is imposed in order  to obtain the asymptotic optimality, which is slightly stronger than that $\xi_n\rightarrow \infty$. Note that Theorem \ref{th-1} holds when both $\lambda_n=2$ and $\lambda_n=\log(n_1)$. These two versions of model averaging methods are both applied in Section \ref{sec4} and Section \ref{sec5}.

\section{Simulation}\label{sec4}
In this section, we conduct a simulation study to compare the finite sample performances of the following methods:

\begin{itemize}
\item CC: a generalized linear regression using subjects that all the covariates are available.
\item SAIC $\&$ SBIC: use the smoothed AIC and smoothed BIC in Buckland et al. (1997) to decide the model weights.
\item IMP: the zero imputation method in  Liu and Zheng (2020). We use IMP1 and IMP2 to denote the IMP method with $\lambda_n=2$ and $\log(n_1)$, respectively.
\item  GLASSO: the method using CC data and group lasso of
Meier et al. (2008) to select covariates and fitting a model
with the subjects that have all the selected covariates available.
\item OPT: the proposed method. We use OPT1 and OPT2 to denote the OPT method with $\lambda_n=2$ and $\log(n_1)$, respectively.
\end{itemize}
The data is generated as follows. A binary $y_i$ is generated from model {\it Binomial}(1,$p_i$) with
$$p_i=\exp(\sum_{j=1}^p\beta_jx_{ij})/\{1+\exp(\sum_{j=1}^p\beta_jx_{ij})\},i=1,\cdots,n,$$
where $p = 14$, $\beta=0.4\times(1,1/2,\cdots,1/p)$, $0.1\times(1,1,\cdots,1)$ or $0.2\times(1/p,\cdots,1/2,1)$,
$x_{i1} = 1$, $(x_{i2},\cdots,x_{ip})$ is generated from a multivariate normal distribution with $E(x_{ij})=1,\text{Var}(x_{ij})=1$, and $\text{Cov}(x_{ij_1},x_{ij_2})=\rho$ for $j_1\neq j_2$, $\rho=0.3$, 0.6 or 0.9, and the sample size $n = 400$ or 800.

\begin{figure}
\centering
\subfloat{%
\resizebox*{14cm}{!}{\includegraphics{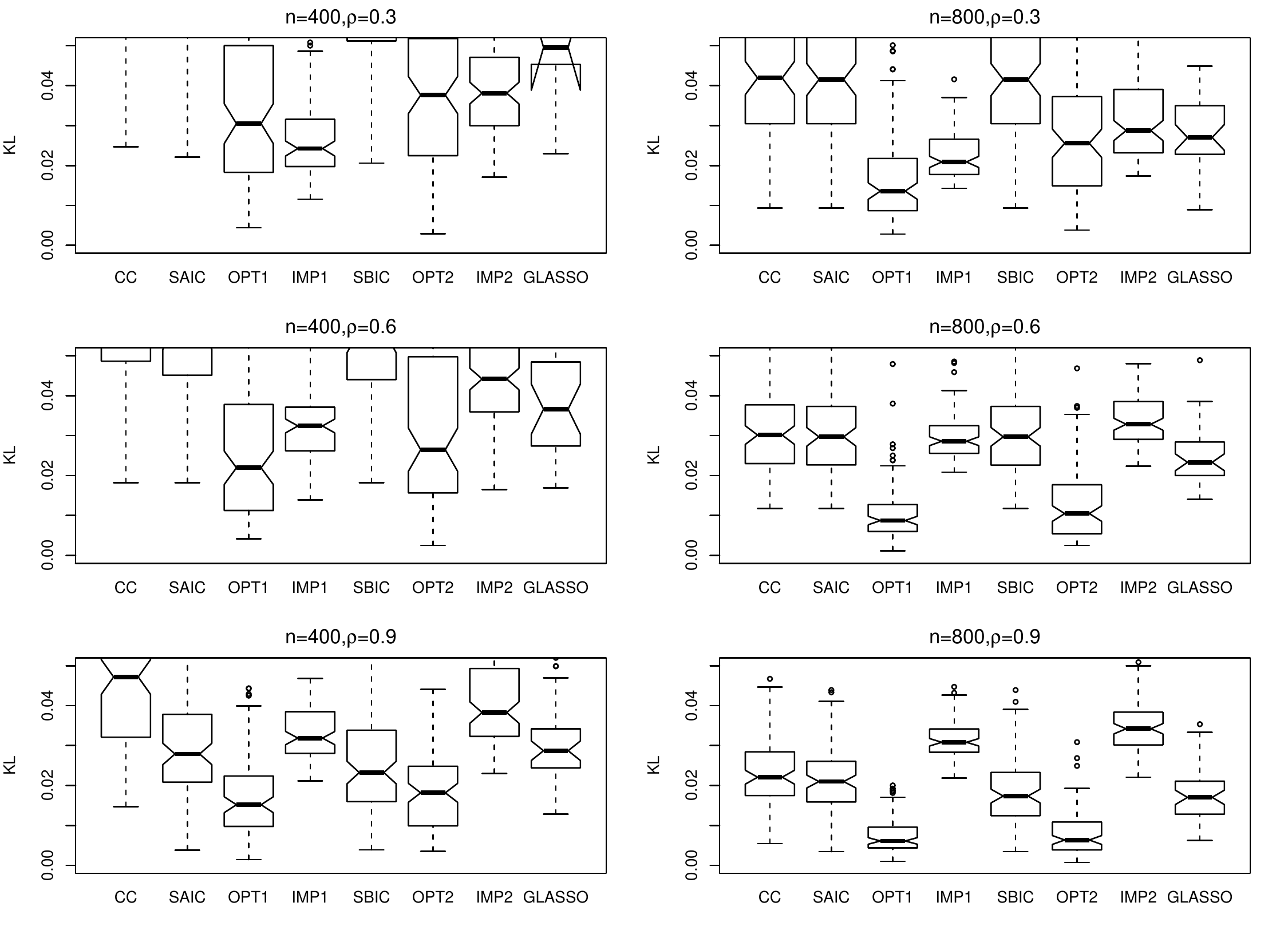}}}\hspace{5pt}
\caption{The KLs of all the methods in 200 replications  when $\beta=0.4\times(1,1/2,\cdots,1/p)$.} \label{tu_1}
\end{figure}
\begin{figure}
\centering
\subfloat{%
\resizebox*{14cm}{!}{\includegraphics{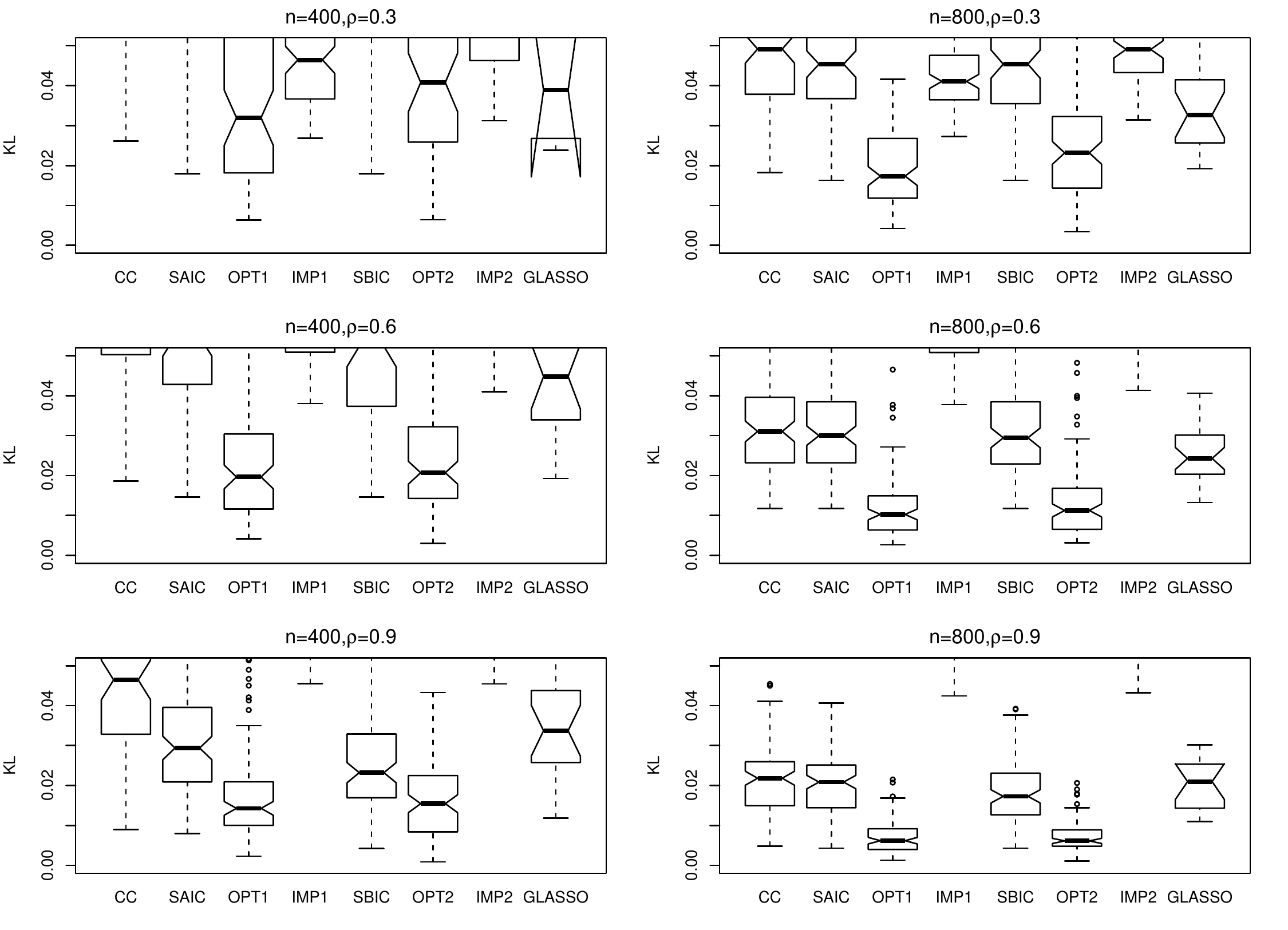}}}\hspace{5pt}
\caption{The KLs of all the methods in 200 replications  when $\beta=0.1\times(1,1,\cdots,1)$.} \label{tu_2}
\end{figure}

\begin{figure}
\centering
\subfloat{%
\resizebox*{14cm}{!}{\includegraphics{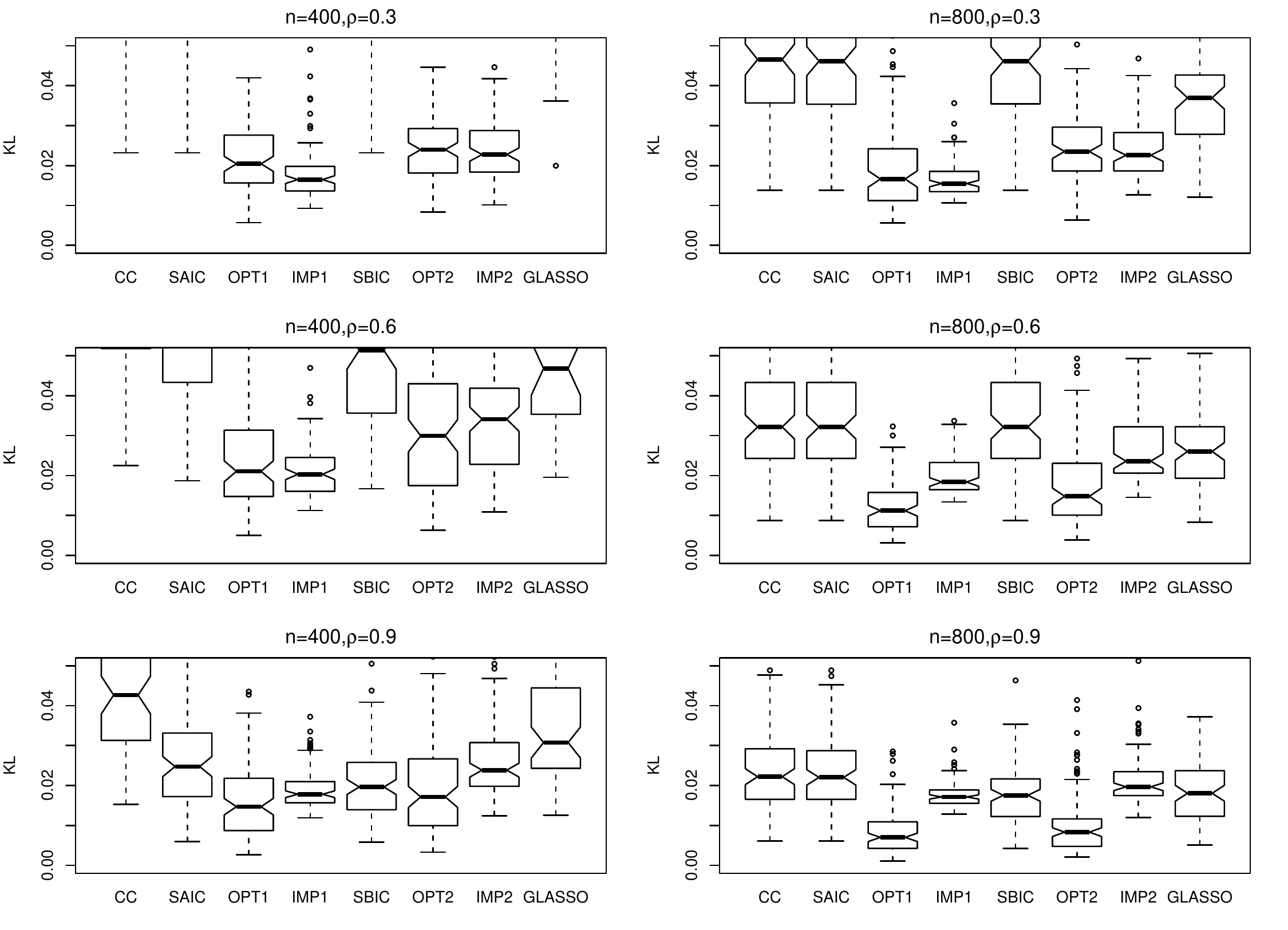}}}\hspace{5pt}
\caption{The KLs of all the methods in 200 replications  when $\beta=0.2\times(1/p,\cdots,1/2,1)$.} \label{tu_3}
\end{figure}

To mimic the situation that all candidate models are misspecified, we pretend that the last covariate is not available for all the candidate models.
The remaining 12 covariates other than the intercept are divided into 3 groups. The $s$th group consists of $X_{4(s-1)+2}$ to $X_{4s+1},$ $s=1,2,3$. The covariates in the sth group are available if the first covariate of each group $X_{4(s-1)+2}<1$,  which results in $K = 8$. The percentages of CC ($S_1$) data are $19\%$, $25.5\%$ and $38.8\%$, respectively for $\rho=0.3$, $0.6$ and 0.9.  We consider the prediction when $D^\ast=D$ and use KL loss (divided by $n_1$) defined in (\ref{KL2}) for assessment. The number of simulation runs is 200. Figures \ref{tu_1} to \ref{tu_3} present the KL loss boxplots for each method  under different simulation settings. The main conclusions are as follows.

(1) The SAIC, SBIC and CC methods perform much worse than OPT1 and OPT2. In many situations, these three methods perform quite similar, indicating that SAIC and SBIC tend to select the model with more covariates and smaller sample size ($M_1$ with CC data).

(2) The zero imputation methods IMP1 and IMP2  generally perform not as well as the proposed methods OPT1 and OPT2. Some exceptions happen when $n$ and $\rho$ are small (for example, the first panel in Figure \ref{tu_1}), in which the usage of zeros to replace unavailable covariates
 has relatively small effect to the prediction.

(3) The performance of GLASSO is also worse than the proposed methods, which shows the model selection method does not work quite well when the models are misspecified.

(4) The proposed method OPT1 produces the lowest KL loss in most situations.

\section{A Real Data Example}\label{sec5}

To illustrate the application of our proposed method, we consider the ADNI  data which is available at {\it http://adni.loni.usc.edu}. The ADNI data contains three different phases: ADNI1, ADNIGO, and ADNI2. In this paper, we use ADNI2 in which some new model data are added. For every subject, different visits at longitudinal time points are recorded and here we focus on the baseline data. As we have mentioned in Section 1, the ADNI data mainly includes four different sources: CSF, PET, MRI and GENE.  The CSF data includes 3 variables: ABETA, TAU and PTAU.  Quantitative variables from the PET images are computed by Helen Wills Neuroscience Institute, UC Berkeley and Lawrence Berkeley National Laboratory containing 241 variables. The MRI is segmented and analyzed in FreeSurfer by the Center for Imaging of Neurodegenerative Diseases at the University of California - San Francisco, which produces 341 variables on volume, surface area, and thickness of regions of interest. GENE, which plays an important role in AD, contains 49386 variables.

\begin{table}
\tbl{Response patterns and sample sizes for ADNI data.}
{\begin{tabular}{c|c|cccc|c}
    \hline
        & \multicolumn{5}{c|}{Data source} &  \\
        \hline
    $k$ & MMSE  & CSF & PET & MRI & GENE & Sample size \\
    \hline
    1   & $\ast$  & $\ast$   & $\ast$   & $\ast$   & $\ast$     & 409 \\
    2  & $\ast$ & $\ast$   & $\ast$   & $\ast$   &       & 368 \\
    3  & $\ast$  & $\ast$   & $\ast$   &     & $\ast$     & 40 \\
    4  & $\ast$ &     & $\ast$   & $\ast$   & $\ast$     & 105 \\
    5  & $\ast$ &     & $\ast$   &     & $\ast$     & 86 \\
    6  & $\ast$ &     & $\ast$   & $\ast$   &       & 53 \\
    7  & $\ast$ &     &     &     & $\ast$     & 53 \\
    8  & $\ast$ &     &     & $\ast$   &       & 56 \\
    \hline
        & &     &     &     & Total & 1170 \\
    \hline
    \end{tabular}}
    \tabnote{{$\ast$}the datum is available.}
\label{table1}%
\end{table}

The overall sample size is 1170. The $K=8$ response patterns and sample size for each pattern are presented in Table \ref{table1}. The total missing rate is about $65\%$. The binary response $Y=1$ if the MMSE score is no less than 28 and $Y=0$ otherwise.

It can be seen that the data is high dimensional, which may contain variables with redundant information. Thus, we first use correlation screening to select features that are most likely to be related to the response variable. All the 3 variables in CSF are kept and 10 variables each for PET, MRI and GENE are screened. We also tried other variable number but found that this screening procedure gave us the smallest KL loss.
\begin{figure}[b]
\centering
\vspace{-0.4cm}
\subfloat{%
\resizebox*{10cm}{!}{\includegraphics{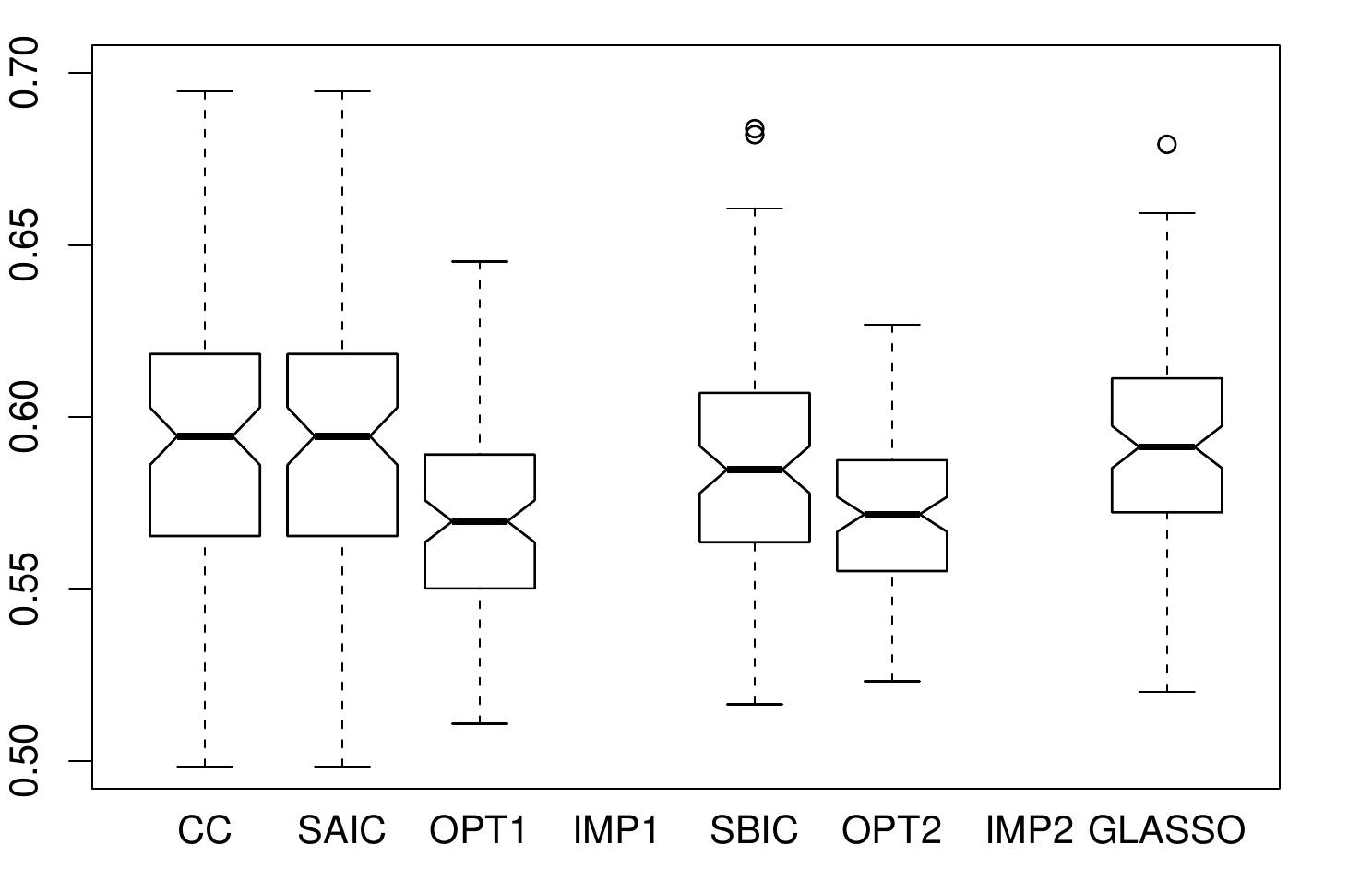}}}\hspace{5pt}
 \vspace{-0.5cm}
\caption{The KL loss of all the methods in 100 replications for ADNI data.} \label{tu_7}
\end{figure}
To compare the prediction performances of the methods considered in the simulation, we randomly select $75\%$ of the subjects from each response pattern and combine them to a training data for model fitting, and use the rest of the subjects as the test data for performance evaluation. For each of the considered methods, we use the training data to fit the model, apply it to the test data, and compute the KL loss of the predictions on test data. We repeat this procedure independently for 100 replications.

Note that in this real data analysis, we do not only consider the prediction for $D^\ast=D$. For $D^\ast \neq\{\text{CSF, PET, MRI, GENE\}}$, the proposed method ignores the covariates not in $D^*$ for modeling and prediction. For example, when $D^*=\{\text{PET, MRI, GENE}\}$, the covariates from ``CSF" are ignored and only 5 candidate models are considered. More details for this kind of procedure can be found in Fang at al. (2019).

Figure \ref{tu_7} displays boxplots of the KL losses over 100 replications for different methods. The boxplots for IMP1 and IMP2 are not shown in the figure because their KL losses are too large. The proposed  methods OPT1 and OPT2 outperform the other methods.

\section{Concluding Remarks}\label{sec6}

Fragmentary data is becoming more and more popular in
many areas and it is not easy to handle. Most existing methods dealing with
fragmentary data consider a continuous response while in many applications the response
variable is discrete. We propose a model averaging method to deal with fragmentary data under generalized linear models. The asymptotic optimality is established and empirical results from a simulation study and a real data analysis about Alzheimer disease show the superiority of the proposed method.

There are several topics for our future study. First, the covariate dimension $p$ and the number of candidate models $K$ are assumed to be fixed. The asymptotic optimality with diverging $p$ and $K$ needs further investigation. Second, we do not focus on the comparison of $\lambda_n=2$  and $\lambda_n=\log(n_1)$. Which tuning parameter should we use in the practice? In fact, how to choose the best tuning parameter for model averaging is still a challenging problem even under linear models. Third, we assume the overall model belongs to an exponential family which is still restrictive. The extension to more general models deserves further study.

\section*{References}
\begin{description}

\item Akaike, H. (1970). Statistical predictor identification. {\it Annals of the Institute of Statistical
Mathematics} {\bf 22}, 203--217.

\item Ando, T. and Li, K.-C. (2014). A model averaging approach for high dimensional regression. {\it Journal of American Statistical Association} {\bf 109}, 254--265.

\item Ando, T. and Li, K.-C. (2017). A weight-relaxed model averaging approach for high-dimensional generalized linear models. {\it The Annals of Statistics} {\bf 45}, 2654--2679.

\item Buckland, S. T., Burnham, K. P. and Augustin, N. H. (1997). Model selection: An integral part of inference. {\it Biometrics} {\bf 53}, 603--618.

\item Chen, J., Li, D., Linton, O. and Lu, Z. (2018). Semiparametric ultra-high dimensional model averaging of nonlinear dynamic time series. {\it Journal of the American Statistical Association} {\bf 113}, 919--932.

\item Dardanoni, V., Luca, G. D., Modica, S. and Peracchi, F. (2015). Model averaging estimation of generalized linear models with imputed covariates. {\it Journal of Econometrics} {\bf 184}, 452--463.

\item Dardanoni, V., Modica, S. and Peracchi, F. (2011). Regression with imputed covariates: A generalized missing indicator approach. {\it Journal of Econometrics} {\bf 162}, 362--368.

\item Ding, X., Xie, J. and Yan, X. (2021). Model averaging for multiple quantile regression with covariates missing at random. {\it Journal of Statistical Computation and Simulation} {\bf 91}, 2249--2275.

\item Fan, J. and Li, R. (2001). Variable selection via nonconcave penalized likelihood and its oracle properties. {\it Journal of American Statistical Association} {\bf 96}, 1348--1360.

\item Fan, J. and Lv, J. (2008). Sure independence screening for ultrahigh dimensional feature space (with discussions). {\it Journal of the Royal Statistical Society, Series B} {\bf 70}, 849--911.

\item Fang, F., Li, J. and Xia, X. (2020). Semiparametric model averaging prediction for dichotomous response. {\it Journal of Econometrics}, In Press.

\item Fang, F., Wei, L., Tong, J., and Shao, J. (2019). Model averaging for prediction with fragmentary data. {\it Journal of Business $\&$ Economic Statistics} {\bf 37}, 517--527.

\item Hansen, B. E. (2007). Least squares model averaging. {\it Econometrica} {\bf 75}, 1175--1189.
				
\item Hansen, B. E. and Racine, J. S. (2012). Jackknife model averaging. {\it Journal of Econometrics} {\bf 167}, 38--46.

\item Hjort, N. L. and Claeskens, G. (2003). Frequentist model average estimators. {\it Journal of American Statistical Association} {\bf 98}, 879--899.

\item Hoeting, J., Madigan, D., Raftery, A. and Volinsky, C. (1999). Bayesian model averaging: A tutorial. {\it Statistical Science} {\bf 14}, 382--401.

\item Kim, J. K. and Shao, J. (2013). {\it Statistical Methods for Handling Incomplete Data}. New York: Chapman \& Hall/CRC.

\item Leung, G. and  Barron, A. R. (2006). Information theory and mixing least-squares regressions. {\it IEEE Transactions on Information Theory} {\bf 52}, 3396--3410.

\item Li, C., Li, Q., Racine, J. S. and Zhang, D. (2018). Optimal model averaging of varying coefficient models. {\it Statistica Sinica} {\bf 28}, 2795--2809.

\item Li, D., Linton, O. and Lu, Z. (2015). A flexible semiparametric forecasting model for time series.
    {\it Journal of Econometrics} {\bf 187}, 345--357.

\item Liao, J., Zong, X., Zhang, X. and Zou, G. (2019). Model averaging based on leave-subject-out cross-validation for vector autoregressions. {\it Journal of Econometrics} {\bf 209}, 35--60.

\item Lin, H., Liu, W. and Lan, W. (2021). Regression analysis with individual-specific patterns of missing covariates. {\it Journal of Business $\&$ Economic Statistics} {\bf 39}, 179--188.

\item Little, R. J. A. and Rubin, D. B. (2002). {\it Statistical Analysis with Missing Data}, 2nd Edition. New York: Wiley.

\item Liu, Q. and Okui, R. (2013). Heteroskedasticity-robust $C_p$ model averaging. {\it The Econometrics
Journal} {\bf 16}, 463--472.

\item Liu, Q. and Zheng, M. (2020). Model averaging for generalized linear model with covariates that are missing completely at random. {\it The Journal of Quantitative Economics} {\bf 11}, 25--40.

\item Longford, N. T. (2005). Editorial: Model selection and efficiency is `Which model... ?' the right question? {\it Journal of the Royal Statistical Society A} {\bf 168}, 469--472.

\item Lu, X. and Su, L. (2015). Jackknife model averaging for quantile regressions. {\it Journal of Econometrics} {\bf 188}, 40--58.

\item Mallows, C. (1973). Some comments on $C_p$. {\it Technometrics} {\bf 15}, 661--675.

\item Meier, L., Geer, S. V. D. and Peter B¨¹hlmann. (2008). The group lasso for logistic regression. {\it Journal of the Royal Statistical Society Series B} {\bf 70}, 53--71.

\item Schomaker, M., Wan, A. T. K. and Heumann, C. (2010).  Frequentist model averaging with missing observations. {\it Computational Statistics and Data Analysis} {\bf 54}, 3336--3347.

\item Schwarz, G. (1978). Estimating the dimension of a model. {\it The Annals of Statistics} {\bf 6},
461--464.

\item Tibshirani, R. (1996). Regression shrinkage and selection via the lasso. {\it Journal of the Royal
Statistical Society, Series B} {\bf 58}, 267--288.

\item Wan, A. T. K., Zhang, X. and Zou, G. (2010). Least squares model averaging by Mallows
criterion. {\it Journal of Econometrics} {\bf 156}, 277--283.

\item White, H. (1982). Maximum likelihood estimation of misspecified models. {\it Econometrica} {\bf 50}, 1--25.

\item Xue, F. and Qu, A. (2021). Integrating multi-source block-wise missing data in model selection. {\it Journal of American Statistical Association} {\bf 116}, 1914--1927.

\item Yang, Y. (2001).  Adaptive regression by mixing. {\it Journal of American Statistical Association} {\bf 96}, 574--588.

\item Yang, Y. (2003). Regression with multiple candidate models: Selecting or mixing? {\it Statistica Sinica} {\bf 13}, 783--809.

\item Zhang, X. (2013). Model averaging with covariates that are missing completely at random. {\it Economics Letters} {\bf 121}, 360--363.

\item Zhang, X., Yu, D., Zou, G. and Liang, H. (2016). Optimal model averaging estimation for generalized linear models and generalized linear mixed-effects models. {\it Journal of the American Statistical Association} {\bf 111}, 1775--1790.

\item Zhang, X., Zou, G. and Liang, H. (2014). Model averaging and weight choice in linear mixed effects models. {\it Biometrika} {\bf 101}, 205-¨C218.

\item Zhang, X., Zou, G., Liang, H. and Carroll, R. J. (2020a). Parsimonious model averaging with a diverging number of parameters. {\it Journal of the American Statistical Association} {\bf 115}, 972--984.

\item Zhang, Y., Tang, N. and Qu, A. (2020b). Imputed factor regression for high-dimensional block-wise missing data. {\it Statistica Sinica} {\bf 30}, 631--651.

\item Zheng, H., Tsui, K-W, Kang, X., and Deng, X. (2017). Cholesky-based model averaging for covariance matrix estimation. {\it Statistical Theory and Related Fields} {\bf 1}, 48--58.

\item Zhu, R., Wan, A. T. K., Zhang, X. and Zou, G. (2019). A Mallow-type model averaging estimator for the varying-coefficient partially linear model. {\it Journal of the American Statistical Association} {\bf 114}, 882--892.
\end{description}

\appendix

\section{Proof of Theorem \ref{th-1}}

\begin{proof}

\renewcommand{\theequation}{A.\arabic{equation}}
			\setcounter{equation}{0}

Let $\tilde{\mathcal{G}}(w)=\mathcal{G}(w)-2\phi^{-1}B_0+2\phi^{-1}\mu_{S_1}^T\theta_0$. It is obvious that $\hat{w}=\argmin_{w\in \mathcal{H}_n}\tilde{\mathcal{G}}(w)$. From the proof of Theorem 1' in Wan et al. (2010), Theorem \ref{th-1} is valid if the following two conclusions hold:
\begin{eqnarray}
  (\text{i}) &&\sup_{w\in \mathcal{H}_n}\frac{|\text{KL}(w)-\text{KL}^*(w)|}{\text{KL}^*(w)}\rightarrow_p0,\label{i}\\
  \text{ and }\hspace{1cm}&&\nonumber\\
  (\text{ii}) && \sup_{w\in \mathcal{H}_n}\frac{|\tilde{\mathcal{G}}(w)-\text{KL}^*(w)|}{\text{KL}^*(w)}\rightarrow_p0. \label{ii}
\end{eqnarray}
By (\ref{con}), we know that uniformly for $w\in \mathcal{H}_n$,
\begin{align}\label{con1}
\hat{\beta}(w)-\beta^*(w)=\sum_{k=1}^Kw_k\Pi_k^T(\hat{\beta}_{(k)}-\beta^*_{(k)})=O_p(n_1^{-1/2}).
\end{align}
It follows from (\ref{con1}), Condition (C1), and Taylor expansion that uniformly for $w\in \mathcal{H}_n$,
$$|B\{\hat{\beta}(w)\}-B\{\beta^*(w)\}|\leq \parallel\frac{\partial B(\beta)}{\partial \beta^T}|_{\beta=\tilde{\beta}(w)}\parallel\parallel\hat{\beta}(w)-\beta^*(w)\parallel=O_p(n_1^{1/2}),$$

$$\mu_{S_1}^T[\theta\{\hat{\beta}(w)\}-\theta\{\beta^*(w)\}]\leq \parallel\mu_{S_1}^T\text{X}_{1}\parallel\parallel\hat{\beta}(w)-\beta^*(w)\parallel=O_p(n_1^{1/2}),$$
and
$$\epsilon_{S_1}^T[\theta\{\hat{\beta}(w)\}-\theta\{\beta^*(w)\}]\leq \parallel\epsilon_{S_1}^T\text{X}_{1}\parallel\parallel\hat{\beta}(w)-\beta^*(w)\parallel=O_p(1),$$
where $\tilde{\beta}(w)$ is a vector between $\hat{\beta}(w)$ and $\beta^*(w)$. In addition, using the central limit theorem and Condition (C.2), we know that uniformly for $w\in \mathcal{H}_n$,
\begin{align}
\epsilon_{S_1}^T\theta\{\beta^*(w)\}=\sum_{k=1}^Kw_k\epsilon_{S_1}^T\theta(\Pi_k^T\beta_{(k)}^*)=O_p(n_1^{1/2}).\nonumber
\end{align}
Then we have
\begin{align}\label{p1}
&\sup_{w\in \mathcal{H}_n}|\text{KL}(w)-\text{KL}^*(w)|\nonumber\\
&\leq 2\phi^{-1}\sup_{w\in \mathcal{H}_n}|B\{\hat{\beta}(w)\}-B\{\beta^*(w)\}|+2\phi^{-1}\sup_{w\in \mathcal{H}_n}|\mu_{S_1}^T[\theta\{\hat{\beta}(w)\}-\theta\{\beta^*(w)\}]|\nonumber\\
&=O_p(n_1^{1/2})
\end{align}
and
\begin{align}\label{p2}
&\sup_{w\in \mathcal{H}_n}|\tilde{\mathcal{G}}(w)-\text{KL}^*(w)|\nonumber\\
&\leq 2\phi^{-1}\sup_{w\in \mathcal{H}_n}|B\{\hat{\beta}(w)\}-B\{\beta^*(w)\}|\nonumber\\
&\quad+2\phi^{-1}\sup_{w\in \mathcal{H}_n}|y^T_{S_1}\theta\{\hat{\beta}(w)\}-\mu_{S_1}^T\theta\{\beta^*(w)\}|+\lambda_n\sum_{k=1}^Kw_kp_k\nonumber\\
&\leq 2\phi^{-1}\sup_{w\in \mathcal{H}_n}|B\{\hat{\beta}(w)\}-B\{\beta^*(w)\}|\nonumber\\
&\quad+2\phi^{-1}\sup_{w\in \mathcal{H}_n}|\mu_{S_1}^T[\theta\{\hat{\beta}(w)\}-\theta\{\beta^*(w)\}]|+2\phi^{-1}\sup_{w\in \mathcal{H}_n}|\epsilon_{S_1}^T\theta\{\beta^*(w)\}|\nonumber\\
&\quad+2\phi^{-1}\sup_{w\in \mathcal{H}_n}|\epsilon_{S_1}^T[\theta\{\hat{\beta}(w)\}-\theta\{\beta^*(w)\}]|+\lambda_n\sum_{k=1}^Kw_kp_k\nonumber\\
&=O_p(n_1^{1/2})+\lambda_n\sum_{k=1}^Kw_kp_k.
\end{align}
Now, from (\ref{p1})-(\ref{p2}), $n_1\xi_n^{-2}=o(1)$, and $n_1^{-1/2}\lambda_n=O(1)$, we can obtain (\ref{i}) and (\ref{ii}). This completes the proof.

\end{proof}
\end{document}